# Stimulated Oscillations in Renewable Energy Integrated Power Systems —Part I: Mechanism and Analysis Methods

Peng Zhang, *Member, IEEE*

***Abstract*--Oscillation is a critical issue that power systems have long faced. Especially over the past two decades, with the large-scale integration of renewable energy into the grid, oscillation problems have posed a serious threat to the secure operation of power systems. However, the current literature has not fully explained the oscillation mechanism of renewable energy integrated power systems (REIPSs). In this paper, the underlying mechanism of stimulated oscillations is explored, with novel analytical methods proposed. Firstly, it is explained from both mathematical formulas and physical interpretations that for an oscillation mode characterized by a pair of complex conjugate poles, the oscillation risk under disturbance depends on the relative positional relationship between the corresponding poles and all other poles and zeros on the complex plane, rather than their standalone locations, i.e., the stability perceived by classical theory. Then the underlying mechanism of high amplitude oscillations induced by closely-located poles under even slight disturbance is clarified. On this basis, a theoretical framework for stimulated oscillations applicable to REIPSs, covering its definition, mechanism, and methods, is proposed. Finally, this paper discusses the relationship between the stimulated oscillation theory proposed herein and the classical stability-based theory, revealing that the research findings surpass rather than negate the classical theories.**

## I. Introduction

Oscillation is a critical issue that power systems have long faced[1]-[13]. Especially over the past two decades, with the large-scale integration of renewable energy into the grid, oscillation problems have posed a serious threat to the secure operation of power systems [14]-[20].

Classic stability-based oscillation analysis and suppression methods, despite their various forms such as eigenvalue analysis, the Routh criterion, the Nyquist criterion, etc., essentially analyze and regulate the pole positions corresponding to oscillation modes on the complex plane, i.e., the stability [1]-[30]. However, stability-based analyses have not fully explained the oscillation mechanism of renewable energy integrated power systems (REIPSs), and field recorded data cannot verify the conclusions of theoretical research [15],[20],[31]-[33]. A typical phenomenon unexplainable by conventional stability theory is that oscillations in actual REIPSs often persist for minutes to hours without divergence, staying far below the onset of nonlinear behaviors such as controller amplitude limiting and component saturation [20],[32]-[33].

References [32]-[35] investigate the non-divergent oscillation mechanism in REIPSs based on the forced oscillation theory. This theory indicates that the necessary conditions for the occurrence of forced oscillations are that the amplitude of periodic disturbances is sufficiently large, the frequency of periodic disturbances is close to the forced mode frequency, and the forced mode is poorly damped [32]-[38]. However, in practical REIPSs with actual non-divergent oscillations, components regarded as disturbance sources, such as wind turbine generators, did not generate periodic disturbance signals with high amplitudes [39]. Moreover, in power systems integrated with large-scale renewable energy, the possibility that numerous components simultaneously produce high-amplitude periodic disturbances of identical frequency is rather low. Therefore, the oscillation occurrence conditions indicated by the forced oscillation theory may be overly idealized and highly hypothetical.

Thus far, a crucial fact seems to have been overlooked by all researchers. In conventional power systems (CPSs), disturbances are discrete and instantaneous. Once an oscillation is triggered by an instantaneous disturbance, the system enters the free oscillation stage. In contrast, in REIPSs, disturbances become continuous due to the inherent fluctuation characteristics caused by variations in wind speed and light intensity. Such disturbances are usually sustained and slight. This raises a critical question: can slight random disturbances induce high-amplitude non-divergent oscillations, and if so, by what mechanism?

In this paper, we reveal that for an oscillation mode characterized by a pair of complex conjugate poles, the oscillation risk under disturbance depends on the relative positional relationship between the corresponding poles and all other poles and zeros on the complex plane, rather than their standalone locations, i.e., the stability perceived by classical theory. Firstly, it is explained from both mathematical formulas and physical interpretations how the relative positional relationship between poles and zeros affects the oscillation risk. Then the underlying mechanism of high amplitude oscillations induced by closely-located poles under even slight disturbance is clarified. On this basis, a theoretical framework for stimulated oscillation applicable to REIPSs, covering its definition, mechanism, and

P. Zhang is with the State Key Laboratory of Alternate Electrical Power System with Renewable Energy Sources (North China Electric Power University), Changping District, China (e-mail: hdbjmoonbird@ncepu.edu.cn).

method is proposed. Finally, this paper discusses the relationship between the stimulated oscillation theory proposed herein and the classical stability-based theory, revealing that the research findings surpass rather than negate the classical theories.

This paper is organized as follows. In Section II, the key factors determining oscillation amplitude in REIPSs are analyzed. In Section III, the mechanism and analysis methods of high amplitude oscillations induced by closely separated poles under slight disturbances are explained and elaborated. Section IV proposes the theoretical framework of stimulated oscillation for REIPSs. Finally, this paper discusses the relationship between the stimulated oscillation theory proposed herein and the classical stability-based theory, revealing that the research findings transcend rather than negate the classical theories.

## II. Key Factors Determining Oscillation Amplitude in REIPSs

### A. The Difference in Oscillation Process between CPSs and REIPSs

A key difference in oscillation process between CPSs and REIPSs appears to have been overlooked by all researchers. In order to explore the factors determining the oscillation risk of REIPSs, it is necessary to first trace the onset and subsequent evolution of oscillations. Fig. 1 shows the difference in typical oscillation processes between CPSs and REIPSs.

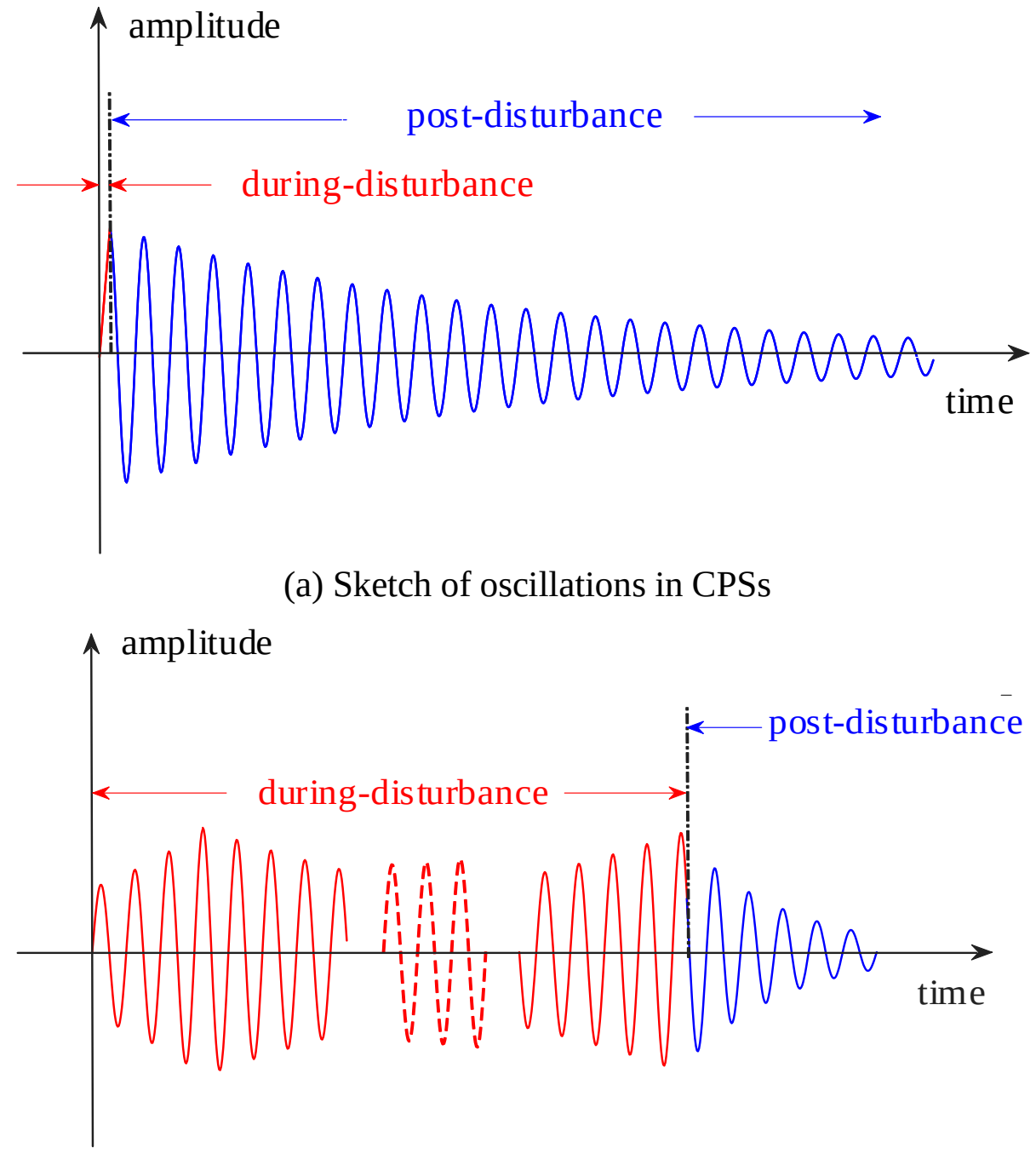

**Fig. 1. Comparison of the oscillation processes of CPSs and REIPSs.**

In CPSs, disturbances are discrete and instantaneous, such as routine switching operations, faults and their removal, load tripping, and other typical transient disturbances. Once an oscillation is triggered by an instantaneous disturbance, the system enters the free oscillation stage. In contrast, in REIPSs, disturbances become continuous due to the inherent fluctuation characteristics caused by variations in wind speed and light intensity. Therefore, in REIPSs the disturbed oscillation process, rather than the free oscillation process determines the oscillation risks.

The classical stability-based analysis methods focus only on the post-disturbance stage. For an oscillatory mode represented by a pair of complex poles σ±jω, the real part σ determines the convergence or divergence of unforced oscillations as well as their decay/growth rates, corresponding to the system dynamics after disturbance. However, disturbances are continuous in REIPSs, and traditional stability-based analysis methods fail to clarify the key factors determining oscillation risk during the disturbance period.

### B. Key Factors Determining Oscillation Amplitude under Disturbances

To reveal the essential mechanism of oscillation, it is necessary to briefly review several fundamental principles. The response of a linear system to a general signal can be solved by decomposing the given signal into a sum of the elementary components and, by superposition, concluded that the response to the general signal is the sum of the responses to the elementary signals.

Consider the general form of the rational function F(s) consisting of the ratio of two polynomials

$$F(s)=\frac{b_1 s^m + b_2 s^{m-1} + \cdots + b_m s + b_{m+1}}{s^n + a_1 s^{n-1} + \cdots + a_n} \quad (1)$$

For any physical system, m<n. By factoring the polynomials, the function can also be expressed in terms of the product of factors as

$$F(s)=K\prod_{j=1}^{m}(s-z_j)\Big/\prod_{i=1}^{n}(s-p_i) \quad (2)$$

Where K is root-locus gain. When s=$z_i$, it is referred to as a zero of the function, and when s=$p_i$, it is referred to as a pole of the function. F(s) can be rewritten via partial fraction expansion

$$F(s)=\frac{c_1}{s-p_1}+\frac{c_2}{s-p_2}+\cdots+\frac{c_n}{s-p_n} \quad (3)$$

The set of coefficients {$C_k$} can be expressed in the following form

$$c_k=(s-p_k)F(s)\big|_{s=p_k} \quad (4)$$

Where $C_k$ is known as the residue of F(s) at pole $p_k$. The coefficient $C_k$ quantifies the contribution of this modal term to the system response.

The partial fraction expansion establishes a direct relationship between disturbance inputs and the corresponding responses of each mode. Accordingly, each partial fraction in the above equation represents the dynamic contribution of the corresponding modal component to the overall response. Hence, the oscillation risk during the disturbance process is governed by the numerator coefficient $C_k$ of each term, rather than the poles in the denominator.

## III. Mechanism and Analysis Methods of High Amplitude Oscillations Induced by Closely Separated Poles under Slight Disturbances

### A. Physical Significance of Coefficient $C_k$

The above brief review of the basic theory indicates that the key factor determining the oscillation amplitude during disturbance is the residue $C_k$. Therefore, it is necessary to investigate the implied physical meanings by combining the calculation method of coefficient $C_k$. Substitute (2) into (4) to obtain

$$C_k = (s-p_k)K\left.\frac{\prod_{j=1}^{m}(s-z_j)}{\prod_{i=1}^{n}(s-p_i)}\right|_{s=p_k} = K\frac{\prod_{j=1}^{m}(p_k-z_j)}{\prod_{i=1,i\neq k}^{n}(p_k-p_i)} \tag{5}$$

On the complex plane, the numerator term, $p_k$-$z_j$, represents the vector from the $k^{th}$ pole to the $i^{th}$ zero. Similarly, the denominator term, $p_k$-$p_i$, represents the vector from the $k^{th}$ pole to the $i^{th}$ pole. Fig. 2 shows the vector schematic of the relative positions between one pole and other poles as well as zeros on the complex plane.

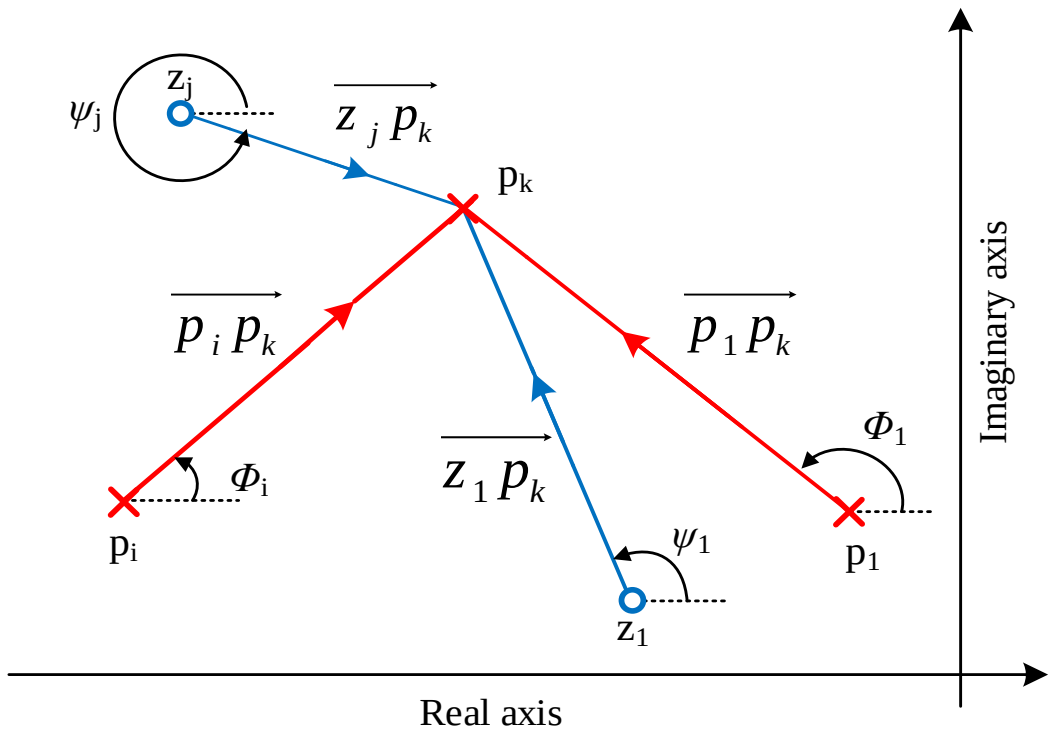

**Fig. 2 Vector schematic of relative positions between one pole and other poles or zeros on the complex plane.**

If $\overrightarrow{z_j p_k}$ represents the vector from $z_j$ to $p_k$ and $\overrightarrow{p_i p_k}$ represents the vector from $p_i$ to $p_k$, as shown in the above figure, (5) can be rewritten as

$$C_k = K\prod_{j=1}^{m}\overrightarrow{z_j p_k}\Big/\prod_{i=1,i\neq k}^{n}\overrightarrow{p_i p_k} \tag{6}$$

Let $\psi_j$ and $\varphi_i$ denote the angles of vectors $\overrightarrow{z_j p_k}$ and $\overrightarrow{p_i p_k}$, respectively; Then (6) can be rewritten as

$$\begin{cases} |c_k| = K\prod_{j=1}^{m}\left|\overrightarrow{z_j p_k}\right|\Big/\prod_{i=1,i\neq k}^{n}\left|\overrightarrow{p_i p_k}\right| \\ \angle c_k = \sum_{j=1}^{m}\psi_j - \sum_{i=1,i\neq k}^{n}\varphi_i \end{cases} \tag{7}$$

Assuming the $(k+1)^{th}$ pole $p_{k+1}$ is the complex conjugate of the $k^{th}$ pole $p_k$, which means $p_k$ and $p_{k+1}$ characterize one oscillation mode, then $C_{k+1}$ is the complex conjugate of $C_k$.

From (6) and (7) it can be found that both the modulus and the angle of $C_k$ is determined by the product of the numerator terms, divided by the product of the denominator terms. When other poles, $p_i$, approach a reference pole, $p_k$, the modulus of the corresponding residue $C_k$ increases, and decreases otherwise. Similarly, when zeros, $z_j$, approach a reference pole, $p_k$, the modulus of the corresponding residue $C_k$ decreases, and increases otherwise.

In this context, the physical significance of the coefficient $C_k$ can be explained as follows: The vectors from the location of the $k^{th}$ pole to all the other poles and zeros jointly determined the corresponding coefficient $C_k$. In other words, the oscillation risk under continuous disturbance conditions is determined by the relative positional relationship between poles and zeros, rather than the absolute locations of poles on the complex plane as stated in classical theories.

### *B. Oscillation Risk Induced by Two Pairs of Closely Separated Complex Conjugate Poles*

Next, we further investigate the oscillation risk induced by two pairs of closely-located complex conjugate poles, i.e., two oscillation modes with similar frequency and damping. First, a simplified scenario is considered, where only these two pairs of poles are present in proximity, with no other poles or zeros.

Assume there is one pair of complex conjugate poles, p1 and $\overset{*}{p}_1$, and another pair of complex conjugate poles, p2 and $\overset{*}{p}_2$, where p1 is closely located to p2, and $\overset{*}{p}_1$ is closely located to $\overset{*}{p}_2$, respectively. where the superscript ∗ represents the complex conjugate. According to (6), the residue corresponding to $p_1$ and $p_2$ are formulated respectively as

$$\begin{cases} c_1 = K\Big/\left(\overrightarrow{\overset{*}{p}_1 p_1}\cdot\overrightarrow{\overset{*}{p}_2 p_1}\cdot\overrightarrow{p_2 p_1}\right) \\ c_2 = K\Big/\left(\overrightarrow{\overset{*}{p}_1 p_2}\cdot\overrightarrow{\overset{*}{p}_2 p_2}\cdot\overrightarrow{p_1 p_2}\right) \end{cases} \tag{8}$$

When $p_1$ is sufficiently close in position to $p_2$, and $\overset{*}{p}_1$ is sufficiently close in position to $\overset{*}{p}_2$, it yields

$$\begin{cases} \overrightarrow{\overset{*}{p}_1 p_1} = \overrightarrow{\overset{*}{p}_1 p_2} \\ \overrightarrow{\overset{*}{p}_2 p_1} = \overrightarrow{\overset{*}{p}_2 p_2} \end{cases} \tag{9}$$

Moreover, given that $\overrightarrow{p_2 p_1} = -\overrightarrow{p_1 p_2}$, (8) can be rearranged as

$$c_1 = K\Big/\left(\overrightarrow{\overset{*}{p}_1 p_1}\cdot\overrightarrow{\overset{*}{p}_2 p_1}\cdot\overrightarrow{p_2 p_1}\right) = -c_2 \tag{10}$$

Accordingly, the partial fraction decomposition of the system with two pairs of complex conjugate poles, namely $p_1$ and $\overset{*}{p}_1=\sigma_1\pm j\omega_1$ and $p_2$ and $\overset{*}{p}_2=\sigma_2\pm j\omega_2$, is expressed below

$$\frac{c_1}{s-(\sigma_1+j\omega_1)}+\frac{\overset{*}{c}_1}{s-(\sigma_1-j\omega_1)}+\frac{c_2}{s-(\sigma_2+j\omega_2)}+\frac{\overset{*}{c}_2}{s-(\sigma_2-j\omega_2)} \tag{11}$$

The corresponding original function, i.e., the time-domain solution contains the following components

$$\begin{aligned} & c_1e^{(\sigma_1+j\omega_1)}+\overset{*}{c}_1e^{(\sigma_1-j\omega_1)}+c_2e^{(\sigma_2+j\omega_2)}+\overset{*}{c}_2e^{(\sigma_2-j\omega_2)} \\ & =2|c_1|\left(e^{\sigma_1 t}\cos(\omega_1 t+\theta_1)-e^{\sigma_2 t}\cos(\omega_2 t+\theta_1)\right) \end{aligned} \tag{12}$$

where $\theta_1$ denotes the initial phase of oscillation modes.

It can be seen from (12) that the time-domain response contains two components with highly similar oscillation frequencies, damping coefficients and initial amplitudes, but opposite initial phases. Next, a set of parameters is substituted to provide an intuitive understanding of the oscillation characteristics.

Assume that a transfer function F(s) is expressed in terms of zero-pole- gain form as

$$F(s)=\frac{K}{\left(s-(\sigma_1+j\omega_1)\right)\left(s-(\sigma_1-j\omega_1)\right)\left(s-(\sigma_2+j\omega_2)\right)\left(s-(\sigma_2-j\omega_2)\right)} \quad (13)$$

Where $\sigma_1=-0.15$, $\omega_1=30\times2\pi$, $\sigma_2=-0.12$, $\omega_2=30.1\times2\pi$. Thus, the transfer function involves two distinct oscillatory modes: a 30Hz oscillation with damping factor 0.15 and a 30.1Hz oscillation with damping factor 0.12，which are close to each other on the complex plane. F(s) is normalized to achieve a unity DC gain (i.e., the gain at s=0) using a gain factor of K=1.2624×109.

The unit impulse responses of the 30.0 Hz and 30.1 Hz modes are plotted in Fig. 3(a), while the composite response is presented in Fig. 3(b).

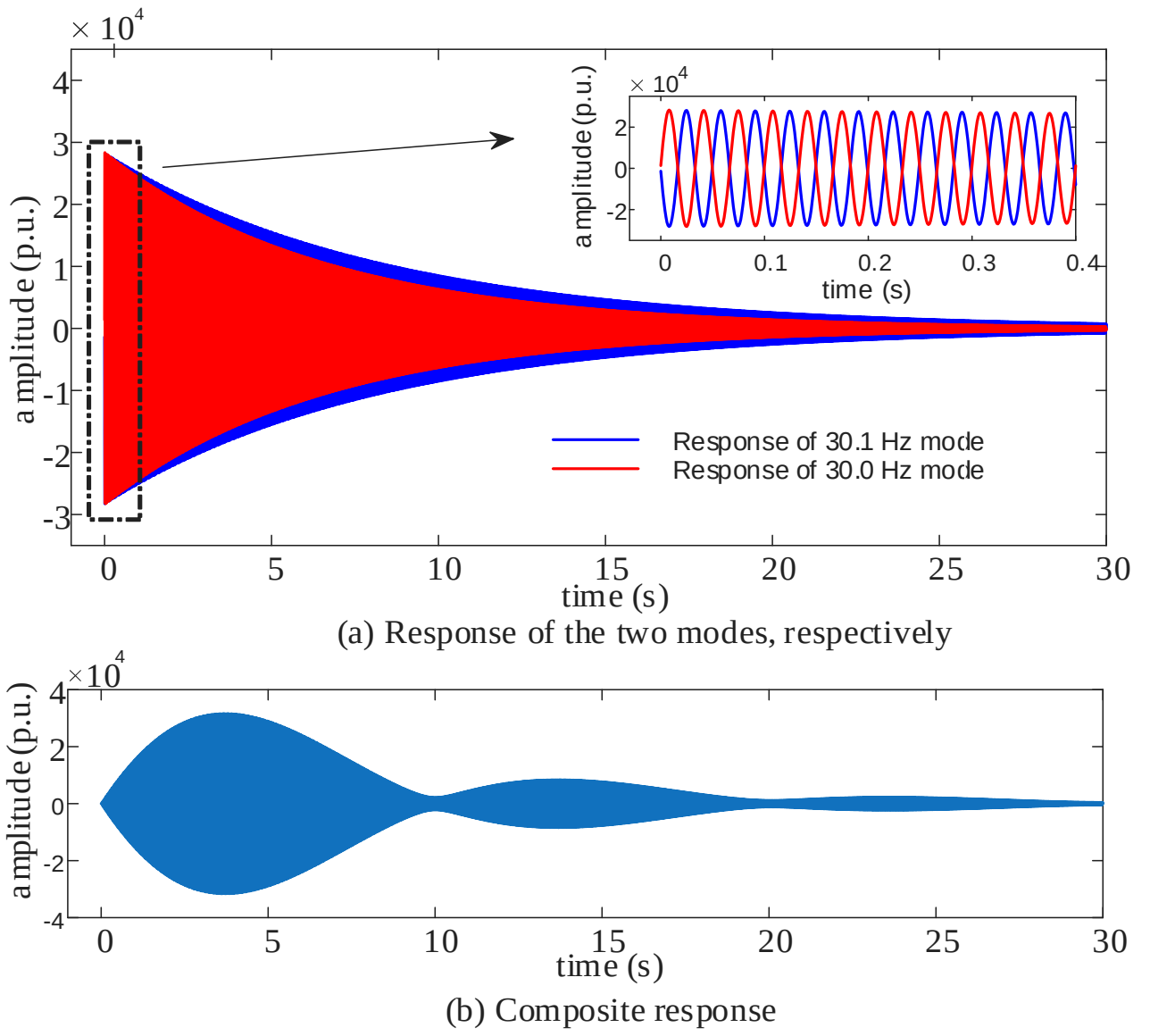

**Fig. 3 Unit impulse responses of two pairs of complex poles close to each other**

Two key conclusions can be drawn from (13) and Fig. 3. Firstly, as long as the distance between two pair of complex poles is sufficiently small, a slight disturbance can induce high oscillation amplitude, regardless of stability. In this case, the DC gain of the transfer function F(s) is 1, however, the amplitude of the unit impulse response can reach the order of $10^4$. Secondly, due to similar oscillation frequencies and opposite initial phases of the two oscillation modes, the oscillation peak does not appear at the initial instant, but rather at the moment of in-phase superposition.

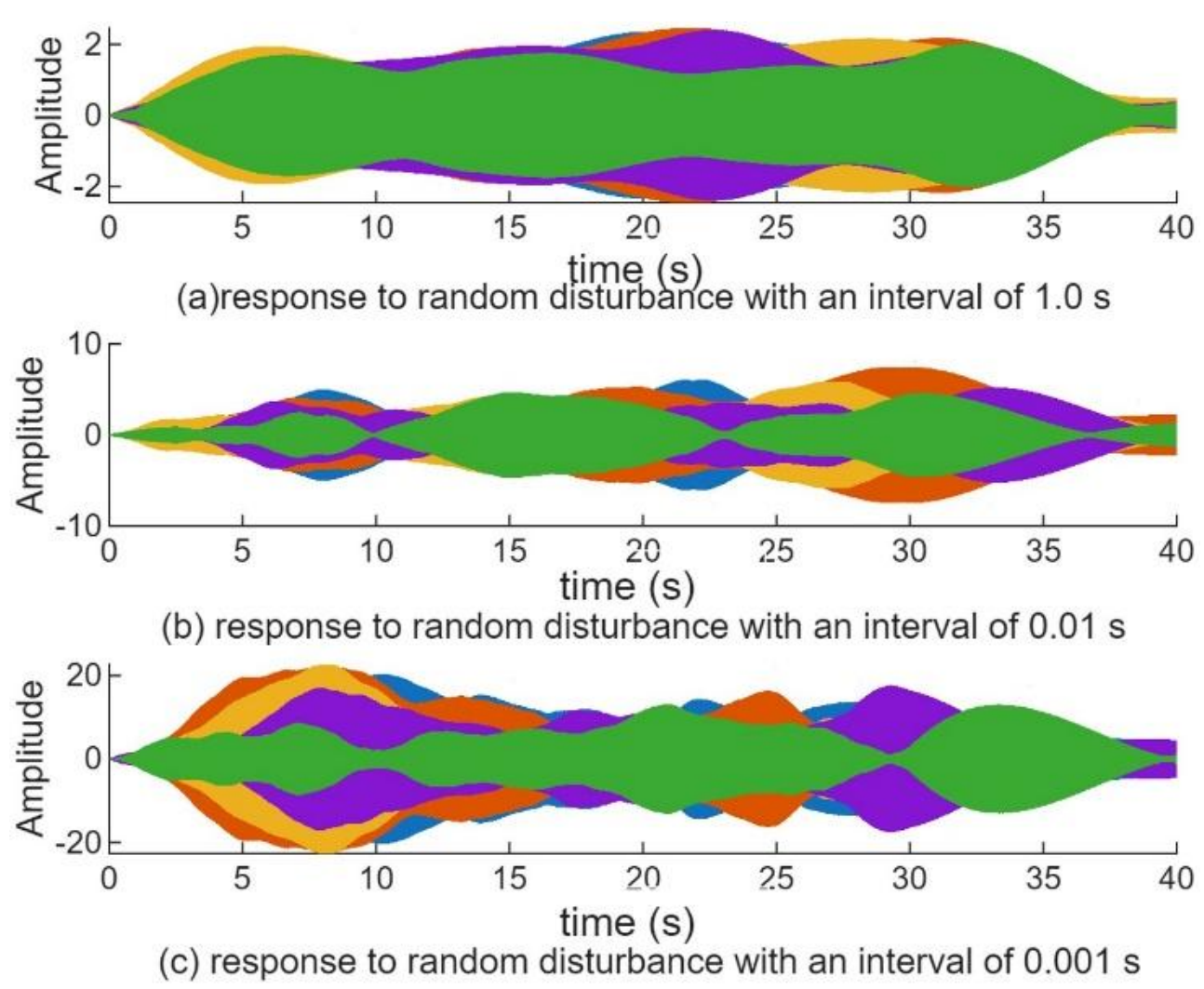

**Fig. 4 Response to continuous random disturbances at different time intervals.**

Low-amplitude persistent random disturbances represent a typical form of disturbances in REIPSs. Under persistent random disturbance conditions, the envelope exhibits irregular fluctuations owing to the constantly changing amplitudes and phases of the two oscillation modes. Fig. 4 shows the system responses to random pulse disturbances with an amplitude range of ±0.01, a duration of 30 s, and pulse intervals of 1 s, 0.01 s, and 0.001 s, respectively，which illustrates the manifestation of high-amplitude oscillations induced by closely separated poles under low-amplitude continuous random disturbances.

### *C. Oscillation Risk of High-Order Systems with Closely Separated Pairs of Complex Conjugate Poles*

Assume high order system F(s) contains *n* poles and *m* zeros, where the $k^{th}$ and $(k+1)^{th}$ poles are a pair of conjugate complex poles, denoted as $p^k$ and $p^{k+1}$, respectively. The residue $C_k$ corresponding to the $k^{th}$ pole is expressed as (6).

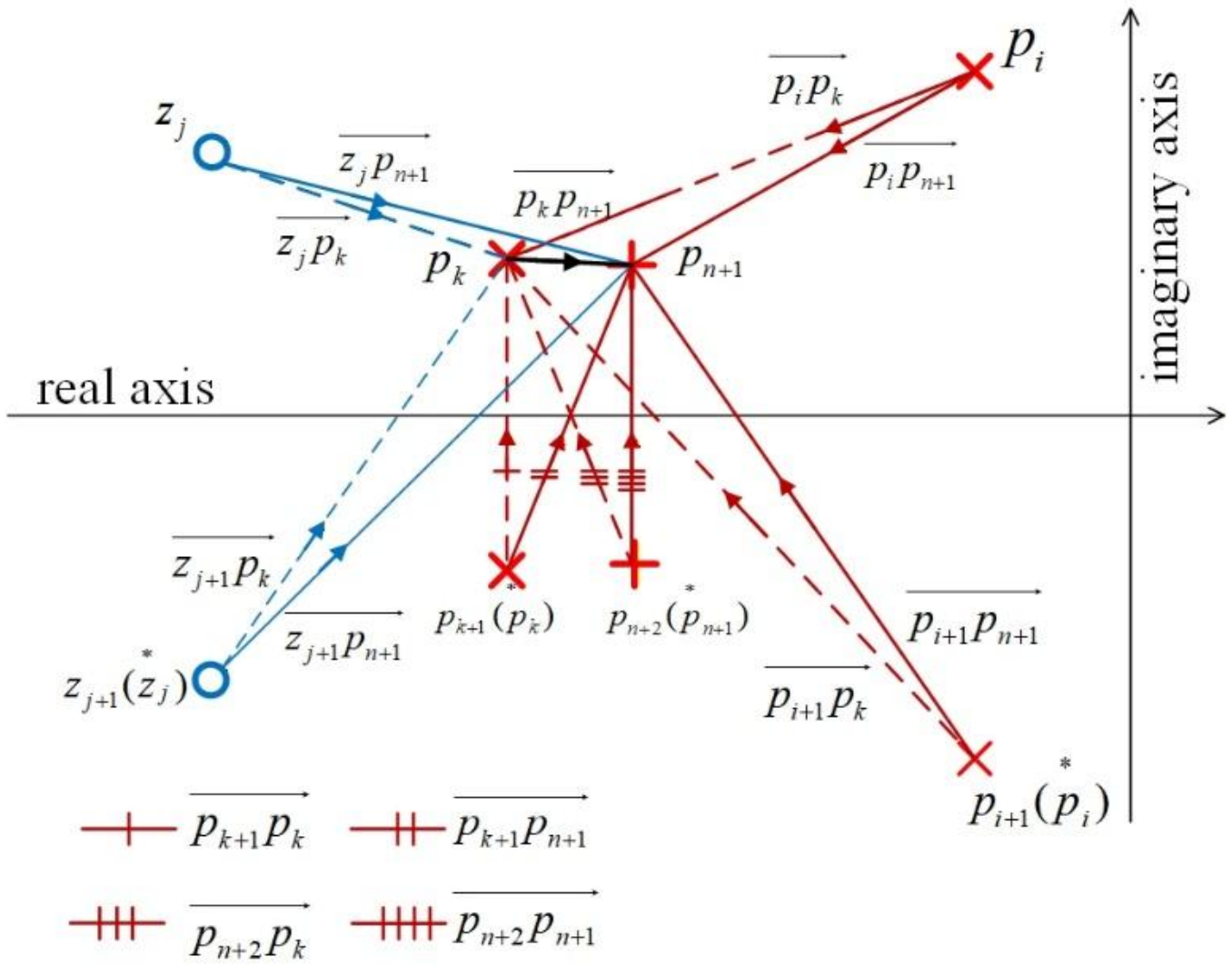

**Fig. 5 Vector schematic of the relative positions for a high-order system with two pairs of poles placed close to each other.**

Now consider adding a new pair of complex conjugate poles, $p_{n+1}$ and $p_{n+2}$, which are placed close to $p_k$ and $p_{k+1}$, respectively. Fig. 5 shows the vector schematic of the relative positions for a high-order system with two pairs of poles placed close to each other.

Then the residue $C_k$ corresponding to the $k^{th}$ pole changes to

$$c_k = K\frac{\prod_{j=1}^{m}\overrightarrow{z_j p_k}}{\prod_{i=1,i\neq k}^{n+2}\overrightarrow{p_i p_k}} = \frac{\prod_{j=1}^{m}\overrightarrow{z_j p_k}}{\prod_{i=1,i\neq k,i\neq n+1}^{n+2}\overrightarrow{p_i p_k}}\frac{K}{\overrightarrow{p_{n+1} p_k}} \tag{14}$$

Similarly, the residue $C_{n+1}$ corresponding to the pole $p_{n+1}$ can be written as

$$c_{n+1} = K\frac{\prod_{j=1}^{m}\overrightarrow{z_j p_{n+1}}}{\prod_{i=1,i\neq n+1}^{n+2}\overrightarrow{p_i p_{n+1}}} = \frac{\prod_{j=1}^{m}\overrightarrow{z_j p_{n+1}}}{\prod_{i=1,i\neq k,i\neq n+1}^{n+2}\overrightarrow{p_i p_{n+1}}}\frac{K}{\overrightarrow{p_k p_{n+1}}} \tag{15}$$

When $p_{n+1}$ is sufficiently close in position to $p_k$, and $p_{n+2}$ is sufficiently close in position to $p_{k+1}$, it yields

$$\begin{cases} \overrightarrow{z_j p_k} = \overrightarrow{z_j p_{n+1}} \\ \overrightarrow{z_{j+1} p_k} = \overrightarrow{z_{j+1} p_{n+1}} \\ \overrightarrow{p_i p_k} = \overrightarrow{p_i p_{n+1}} \\ \overrightarrow{p_{i+1} p_k} = \overrightarrow{p_{i+1} p_{n+1}} \\ \overrightarrow{p_{k+1} p_k} = \overrightarrow{p_{k+1} p_{n+1}} \\ \overrightarrow{p_{n+2} p_k} = \overrightarrow{p_{n+2} p_{n+1}} \\ \overrightarrow{p_k p_{n+1}} = -\overrightarrow{p_{n+1} p_k} \end{cases} \tag{16}$$

Where $z_{j+1}$denotes the conjugate complex $z_j$.

Substitute (16) into (15) to obtain

$$c_{n+1} = -c_K \tag{17}$$

Then, complete partial fraction expansion form of F(s) can be expressed as

$$F(s) = \frac{c_1}{s-p_1} + \cdots + \frac{c_k}{s-p_k} + \frac{c_{k+1}}{s-p_{k+1}} + \cdots + \frac{c_n}{s-p_n} + \frac{c_{n+1}}{s-p_{n+1}} + \frac{c_{n+2}}{s-p_{n+2}} \tag{18}$$

Since the modulus of the vector $\overrightarrow{p_k p_{n+1}}$ and $\overrightarrow{p_{k+1} p_{n+2}}$ are extremely small, the residues, $C_k$, $C_{k+1}$, $C_{n+1}$, and $C_{n+2}$, are significantly larger than the other residues. Accordingly, (18) can be simplified into a form identical to (11), as shown below.

$$F(s) = \frac{c_k}{s-p_k} + \frac{\overset{*}{c_k}}{s-p_{k+1}} + \frac{c_{n+1}}{s-p_{n+1}} + \frac{\overset{*}{c_{n+1}}}{s-p_{n+2}} \tag{19}$$

As can be seen from (19), when a high-order system contains two pairs of closely-located complex conjugate poles, its partial-fraction expansion can be simplified to retain only the terms corresponding to these pairs of poles. In other words, for a complex high-order system with two similar oscillation modes, i.e., when the positions of the two pairs of complex conjugate poles are close to each other, the oscillation risk under sustained disturbances can be characterized solely by the partial-fraction corresponding to these two pairs of poles. This conclusion greatly facilitates the oscillation analysis of high-order systems and provides potential methodological guidance for oscillation suppression. Some recorded waveform data are provided below to corroborate this conclusion.

In a renewable-thermal bundled power system consisting of a large-scale wind farm and two thermal power plants A and B, severe shaft torsional vibration occurred in the thermal generating units of both Plant A and Plant B during an oscillation event on July 1, 2015.

Figs. 6 and 7 present the recorded waveforms and corresponding frequency spectra of the generator speed deviation for the generating units in thermal power plant B from 11:48:21 to 11:49:20.

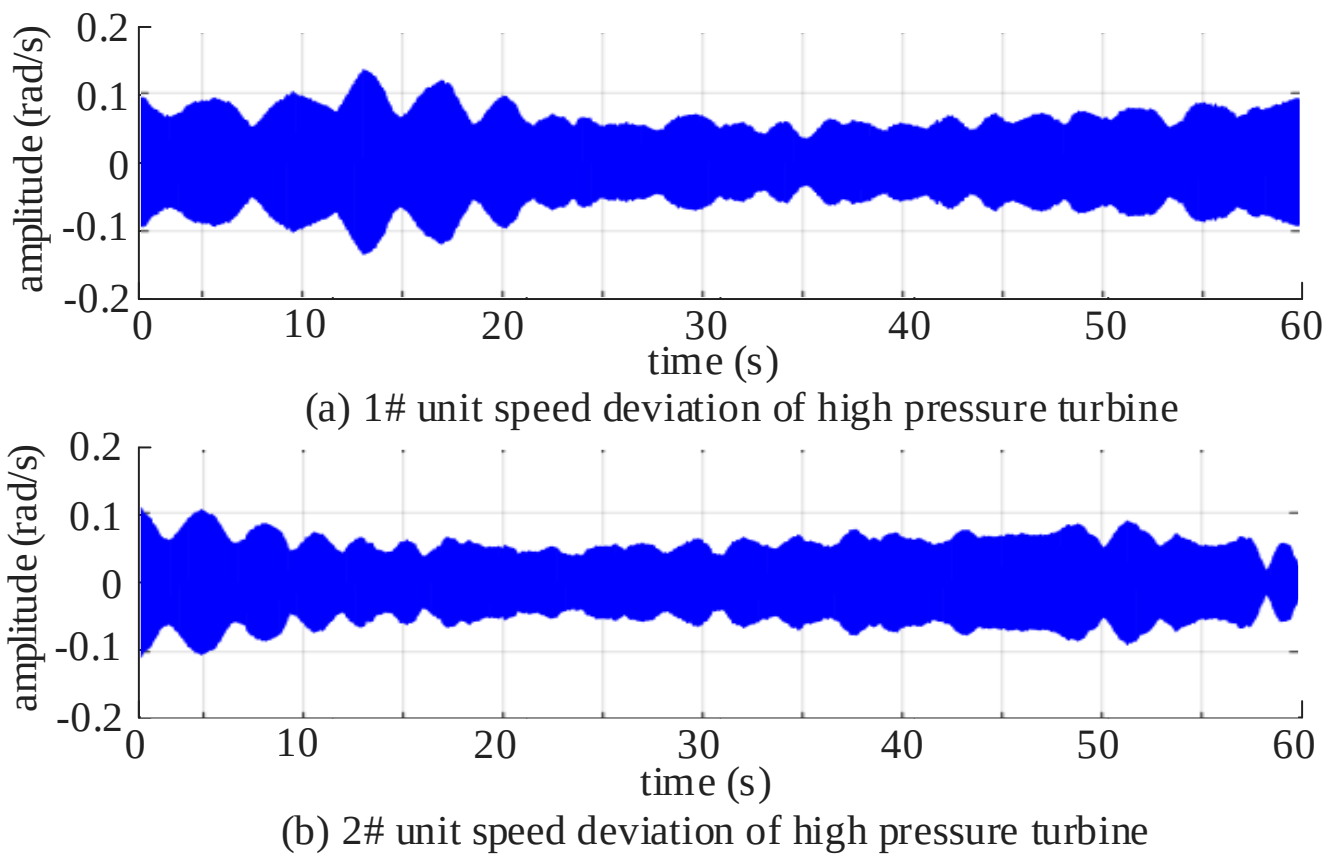

**Fig. 6 field recorded curves of rotor speed deviation in plant B**

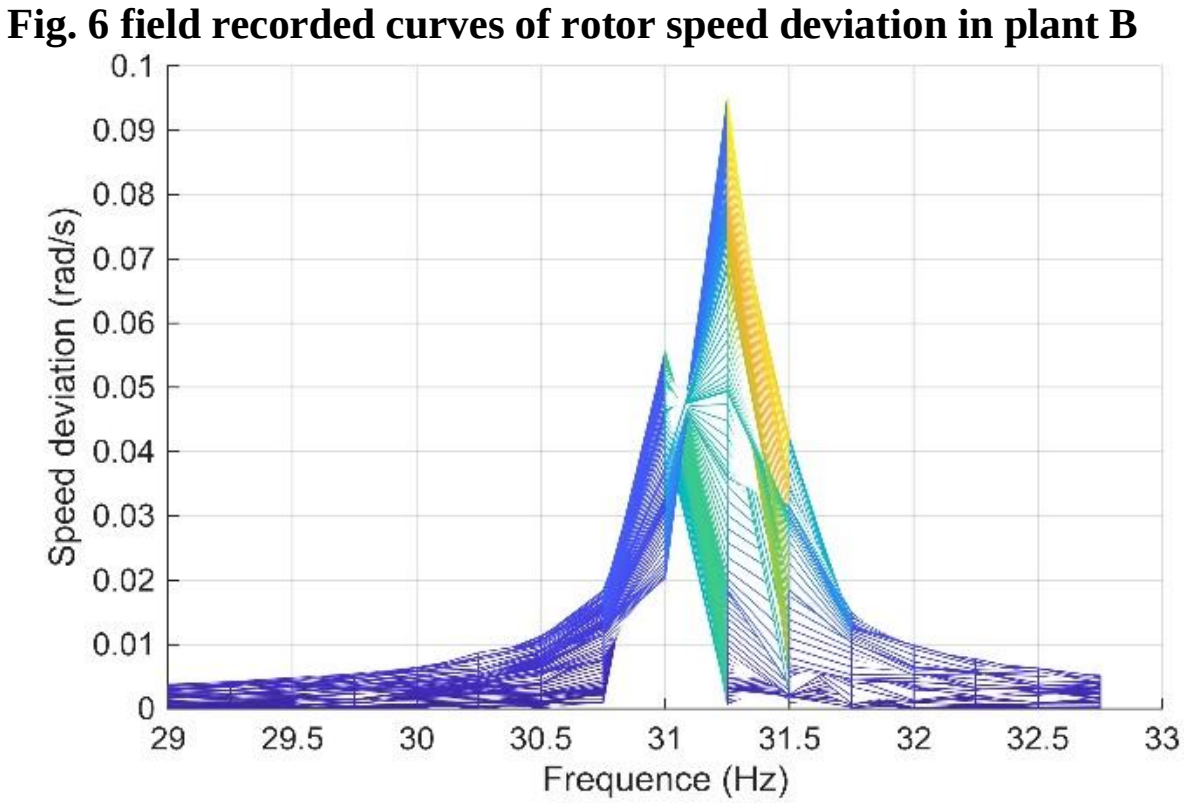

**Fig. 7 Frequency spectrum of the speed deviation curve in plant B**

It can be seen from Figs. 6 and 7 that the rotor speed deviation curve contains exactly two oscillation modes: one is the shaft torsional mode at approximately 31.3 Hz, and the other is the electrical oscillation mode at around 31 Hz.

Figs. 8 and 9 illustrate the recorded waveforms and frequency spectra of the generator speed deviation for the generating units in thermal power plant A during the period of 11:51:20 to 11:52:20.

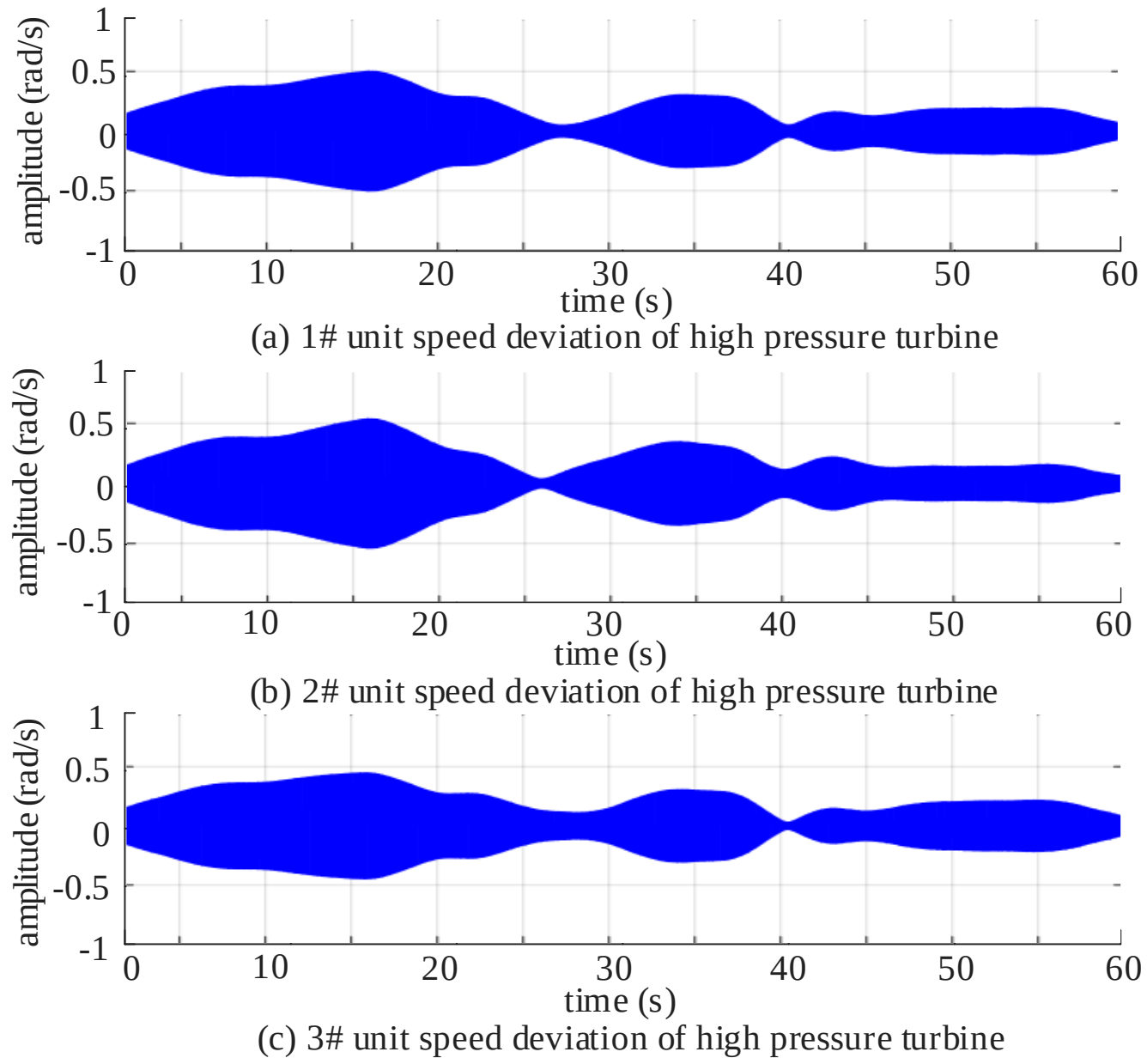

**Fig. 8 field recorded curves of rotor speed deviation in plant A**

**Fig. 9 Frequency spectrum of the speed deviation curve in plant A**

It can be seen from Figs. 8 and 9 that the rotor speed deviation curve contains two oscillation modes: one is the shaft torsional mode at approximately 30.5 Hz, and the other is the electrical oscillation mode at around 30.7 Hz.

## IV. Theoretical Framework of Stimulated Oscillations for REIPSs

### A. Poles Distribution Characteristics of REIPSs

CPSs usually consist of synchronous generators and their controllers, transformers, transmission lines, and loads, all of which belong to low-order systems. For instance, a synchronous generator model incorporating the dynamics of the excitation winding, damping winding, and rotor motion equations is a 7th-order model [6]. Excitation controllers and governors are typically 1st to 3rd-order models [6]. Furthermore, their dynamic characteristics mainly manifest as low-frequency electromechanical oscillations [23]. Accordingly, when the dynamic characteristics of these components are described via poles in the complex plane, the number of poles is relatively small, and these poles are predominantly distributed in the regions close to the real axis. Thus, for CPSs, it is rare to see poles in close proximity to each other.

On the contrary, the probability of poles being closely separated rises significantly for REIPSs. Particularly over the past two decades, profound transformations have taken place in modern power systems, accompanied by the extensive integration of power electronic devices. Representative technologies include FACTS, conventional HVDC, and MMC-HVDC, with wind and photovoltaic power generation being the most typical applications.

Power electronic devices are typically high-order systems [16]-[17], [19]-[20], [24], [27]-[30]. For instance, a model that incorporates not only the detailed control of the grid-side converter, but also the control of the machine-side converter and type-4 generator dynamics results in a 47th-order state-space model [40]. Similarly, an MMC-HVDC inverter can be represented as a 16th-order model [41].

As the quantity and variety of power electronic devices continue to grow, the probability that their characteristic poles approach one another or overlap with those of other devices increases significantly on the complex plane.

### B. Mechanism of Stimulated Oscillations in REIPSs

Accordingly, it is essential to summarize and consolidate the key conclusions drawn from the above analysis to clarify the stimulated oscillation mechanism.

Firstly, in terms of the oscillation evolution, the inherent fluctuating nature of renewable energy excites disturbed oscillations instead of free oscillations, thereby determining the essential oscillation characteristics of the system. Traditional stability-oriented oscillation theories are limited to analyzing free oscillation risks. Accordingly, they apply only to CPSs dominated by free oscillations, and are not suitable for REIPSs where disturbed oscillations prevail.

Secondly, under persistent disturbances, the oscillation risk of a specific oscillation mode is determined by the relative positions on the complex plane between the poles corresponding to this mode and all other poles and zeros, rather than by the standalone location of its own pole as claimed in classical stability-based oscillation theories.

Thirdly, when the locations of two pairs of complex conjugate poles are in close proximity, i.e., when the frequency and damping of the two modes are similar, slight disturbances can induce high amplitude oscillation. Furthermore, for a complex high-order system with two pairs of closely-located complex conjugate poles, the oscillation risk under sustained disturbances can be characterized solely by the partial-fraction corresponding to these two pairs of poles.

Fourthly, the extensive integration of power electronic devices significantly increases the probability that their characteristic poles approach one another or coincide with those of other devices. Furthermore, the variability of renewable energy operation modes causes the positions of closely spaced poles on the complex plane to vary over a wide range. That is, the frequency at which high-amplitude oscillations occur drifts across a broad frequency band as the operating conditions of renewable energy sources change.

Finally, disturbances that excite oscillations can be general or random signals, rather than requiring specific excitation signals, e.g., high-amplitude and/or periodic signals as indicated by the forced oscillation theory.

### C. Theoretical Framework for Stimulated Oscillations

Accordingly, this study tentatively proposes a theoretical framework for stimulated oscillation applicable to REIPSs, covering its definition, mechanism, and methods.

Definition: Stimulated oscillation is phenomenon that high amplitude oscillations are induced by slight disturbances when two oscillation modes, although both of which are stable, are highly similar.

Mechanism: When two oscillation modes are highly similar, their response components also contribute significantly to the total response.

Method: for a high-order system with two similar oscillation modes, the oscillation risk under sustained disturbances can be characterized solely by the partial-fraction corresponding to these two pairs of poles corresponding to the similar oscillation modes.

The theoretical framework proposed above is only preliminary. We maintain an open and sharing attitude and hope that it can be examined and improved by academic peers, so as to gradually form a widely recognized and instructive theoretical system.

## V. Discussion

### A. The Relationship between the Stimulated Oscillation Theory Proposed Herein and Classical Stability-Based Theories

The stimulated oscillation theory proposed in this paper is motivated by a critical fact that the disturbed oscillation process, rather than the free oscillation process, dominates the oscillation behavior in REIPSs. Under this condition, the relative positional relationship between the poles and zeros, rather than the standalone locations of poles, i.e., the stability, determines the oscillation risk. This outcome is fundamentally caused by the grid integration of massive power electronic devices and the inherent random fluctuation characteristics of renewable energy. In spite of this, for CPSs and REIPSs, stability remains a fundamental prerequisite for secure power system operation. The research findings surpass rather than negate the classical theories.

Although not addressed in this work, the risk of instability oscillations in the conventional sense has in fact also increased for REIPSs. Both instability oscillations and stimulated oscillations are concurrent challenges faced by REIPSs. Accordingly, oscillation analysis and mitigation should be considered from both perspectives, which could pose a new challenge for future research and industrial applications.

### B. Methodology for Stimulated Oscillation Mitigation

The mechanism analysis of stimulated oscillations presented in this paper also provides methodological insights for oscillation mitigation. Regardless of the control strategies adopted or the equipment implemented, the overall objective is to separate closely approaching poles and/or bring zeros close to neighboring poles. This, of course, is under the premise that all poles remain within the half-plane of the complex plane.

Since the mechanisms and mitigation methods of stimulated oscillation and unstable oscillation are fundamentally different, it is reasonable to infer—without rigorous theoretical proof—that applying stability-based approaches to suppress stimulated oscillation may be counterproductive. Therefore, identifying the type of oscillation risk is a prerequisite.

It is difficult to predict oscillation risks in the absence of accurate parameters for equipment such as wind turbines. For engineering practice, identifying the type of oscillation risk based on the distinct characteristics of stimulated oscillations and instability oscillations may be a feasible approach. For instance, the envelope of a free oscillation curve follows an exponential law, and the second derivative of an exponential function is always positive. This implies that regardless of whether the first derivative of the exponential function is positive or negative—that is, whether the free oscillation is convergent or divergent—the upper envelope is always concave downward. However, the envelope of stimulated oscillations does not follow this rule.

## VII. Biographies

**Peng Zhang** (M’2010) received his B.S. and Ph.D degree in electrical engineering from North China Electric Power University, in 1999 and 2014. He joined the School of Electrical and Electronic Engineering, North China Electric Power University, in Mar. 2005. His research interests lie in power system analysis and control, with a focus on power system oscillations.